\documentclass[fleqn,usenatbib]{mnras}

\usepackage[T2A,T1]{fontenc}
\usepackage[utf8]{inputenc}
\usepackage{ae,aecompl}

\usepackage{graphicx}	% Including figure files
\usepackage{amsmath}	% Advanced maths commands
\usepackage{amssymb}	% Extra maths symbols

\title[Periodic fast radio bursts]{Periodic fast radio bursts from forcedly precessing neutron stars, anomalous torque, and internal magnetic field for FRB 180916.J0158+65 and FRB 121102}

\author[D. N. Sob'yanin]{
Denis Nikolaevich Sob'yanin\thanks{E-mail: sobyanin@lpi.ru}
\fontencoding{T2A}\selectfont
 (Денис Николаевич Собьянин)
\fontencoding{T1}\selectfont
\\
P. N. Lebedev Physical Institute of the Russian Academy of Sciences, Moscow 119991, Russia
}

\date{Accepted 2020 July 2. Received 2020 June 30; in original form 2020 May 27}

\pubyear{2020}

\begin{document}
\label{firstpage}
\pagerange{\pageref{firstpage}--\pageref{lastpage}}
\maketitle

% Abstract of the paper
\begin{abstract}
A recent discovery of the periodic activity of the repeating fast radio burst source FRB 180916.J0158+65 in the Canadian Hydrogen Intensity Mapping Experiment (CHIME) hints at possible origin of the FRB from a freely precessing neutron star with a magnetar magnetic field of about $10^{16}$~G. However, the absence of simultaneously detected high-energy emission in the Swift and AGILE observations imposes stringent constraints on the field magnitude and questions the possibility of such a progenitor. We show that consideration of forced precession of a neutron star does not encounter the difficulty. This kind of precession takes place even if the neutron star is not deformed and is brought about by the anomalous moment of electromagnetic forces induced by stellar rotation and determined by non-corotational currents. Contrary to what is expected for the currents of corotation, the anomalous torque calculated by the direct method appears to be non-zero. If the observed 16.35-day period corresponds to the period of stellar precession, the inferred internal magnetic field appears to be about $6\times10^{14}$~G for rotational period 1~s. For another possibly periodic FRB~121102 with 157-day period the magnetic field is even lower, $2\times10^{14}$~G, thereby justifying earlier considerations and not ruling out the hypothesis of FRB origin from precessing neutron stars.
\end{abstract}

% Select between one and six entries from the list of approved keywords.
% Don't make up new ones.

%%%%%%%%%%%%%%%%%%%%%%%%%%%%%%%%%%%%%%%%%%%%%%%%%%

%%%%%%%%%%%%%%%%% BODY OF PAPER %%%%%%%%%%%%%%%%%%
\section{Introduction}

Fast radio bursts (FRBs) are cosmic transients revealing themselves as solitary bursts of millisecond duration \citep{PetroffHesselsLorimer2019,CordesChatterjee2019}. High dispersion measures and association of 5 sources with host galaxies (redshift range 0.03-0.66) imply cosmological distances to FRBs \citep{TendulkarEtal2017,BannisterEtal2019,RaviEtal2019,ProchaskaEtal2019,MarcoteEtal2020}. At present, the bursts are detected exclusively in radio, usually at frequencies $\sim1$~GHz with typical energy flux $\sim1$~Jy and total energy release on the assumption of isotropy $\sim10^{40}$~erg. In 2016, the first repeating FRB activity was discovered \citep{SpitlerEtal2016}, pertaining to the earlier known FRB 121102, and over the past year the number of repeating sources is significantly increased, now amounting to 20 \citep{AndersenEtal2019,KumarEtal2019,FonsecaEtal2020}.

FRB 180916.J0158+65 was first detected in the Canadian Hydrogen Intensity Mapping Experiment (CHIME) through 10 bursts in the frequency band 400--800~MHz with the peak flux lying in the range 0.4--4~Jy as a radio source analogous to other repeating FRBs \citep{AndersenEtal2019}. Subsequent VLBI observations allowed one to associate FRB~180916 with a star-forming region in the massive spiral hosting galaxy SDSS J015800.28+654253.0 located at a distance of about 150~Mpc from us, making the FRB the closest among the presently detected (redshift $z\approx0.034$) \citep{MarcoteEtal2020}. FRB~180916 attracted immense interest after the discovery of an unexpected regular activity characterized by the existence of a periodic 16-day modulation: The observed radio bursts appear to be distributed not randomly but are concentrated in 5-day time intervals around epochs of peak activity without detection beyond the intervals, and the epochs repeat every 16.35~days \citep{AmiriEtal2020}. Very recently, the FRB was also detected at even lower radio frequences, 3 bursts at 328~MHz with the Sardinia Radio Telescope \citep{PiliaEtal2020} and 7 bursts in the frequency range 300--400~MHz with the Green Bank Telescope \citep{ChawlaEtal2020}. In addition, information about possible detection of periodicity from the already mentioned first repeater FRB~121102 with the Lowell Telescope was appeared, and the most likely period $\sim157$~days was proposed \citep{RajwadeEtal2020}.

The nature of single, repeating, or periodic FRBs is not established unambiguously so far, and there were proposed numerous theoretical models relating FRBs to neutron stars, black holes, or more exotic objects \citep{PopovPostnov2013,PenConnor2015,LieblingPalenzuela2016,MuraseKashiyamaMeszaros2016,MetzgerBergerMargalit2017,Istomin2018,YangDai2019,
ConnorMillerGardenier2020,IokaZhang2020,BeniaminiWadiasinghMetzger2020,ChernoffLiTye2020,DaiZhong2020,LyutikovBarkovGiannios2020,
LyutikovPopov2020,YangZou2020}. The bursting periodicity of FRBs naturally implies some underlying periodic physical phenomenon. For example, tilted accretion discs around rotating black holes experience Lense-Thirring precession leading to concomitant precession of the generated relativistic jets, and an indication of this scenario is seemingly observed for the M87 jet as its quasi-periodic sideways shift \citep{Sobyanin2018}. This idea was used to explain the observed periodicity of FRB~180916 as being related to jet precession \citep{Katz2020}. An alternative explanation draws an idea of free precession of a neutron star the magnetic field of which is strong enough to cause stellar deformation and may even reach $\sim10^{16}$~G, which hints at the possible magnetar nature of the FRB \citep{LevinBeloborodovBransgrove2020,ZanazziLai2020}. This mechanism of the appearance of free precession was invoked earlier to explain the observed slow phase modulation of periodic pulsations in hard X-rays from the magnetar 4U~0142+61 and gave the same magnitude of the magnetic field \citep{MakishimaEtal2014}.

Meanwhile, \citet{TavaniEtal2020} have studied the Swift and AGILE observations of FRB 180916 with a special focus on the 5-day time intervals in which the FRB is active in radio and also searched any temporal coincidence at millisecond time scales between the known radio bursts and possible high-energy events. The large energy output of the FRB in radio naturally implied possible concomitant high-energy emission, which would somehow clarify the nature of the FRB itself and the physical phenomenon underlying its unusual periodic activity. Quite unexpectedly, there was no detection of any simultaneous event or any extended X-ray and $\gamma$-ray emission on time scales of hours, days, or weeks. The negative result allowed the authors to obtain upper limits on the high-energy emission from this source, $\sim1.5\times10^{41}\text{ erg}\,\text{s}^{-1}$ for isotropic X-ray luminosity  (range $0.3-10$~keV) and the flux an order of magnitude higher for average gamma luminosity (range $>100$~MeV).

On the basis of these energy limits, \citet{TavaniEtal2020} derived a constraint applicable to a neutron star of magnetar type that might be responsible for the observed bursting activity of the periodic FRB,
\begin{equation}
\label{RBtau}
\Bigl(\frac{R}{10\text{ km}}\Bigr)^3\Bigl(\frac{B}{10^{16}\text{ G}}\Bigr)^2\Bigl(\frac{\tau}{10^8\text{ s}}\Bigr)^{-1}\lesssim1,
\end{equation}
where $R$ is the stellar radius, $B$ is the internal magnetic field, and $\tau$ is the characteristic time of magnetic dissipation. From relation \eqref{RBtau} the authors concluded that the values of $\tau$ in the order of the time scales of the conducted observations are generally incompatible with the existence of a high magnetic field of $10^{16}$~G in the magnetar-like progenitor of FRB~180916. Only a magnetar with extreme properties in terms of its magnetic field or dissipation process might power the observed radio bursts. After a millisecond radio burst the hypothetic magnetar should experience overall relaxation on a much longer time scale.

In this paper we intend to show that the above clash of the theoretical hypothesis of the magnetar origin of periodic FRBs and the present results of radio and high-energy observations does not emerge when considering forced precession of a strongly magnetized neutron star. Such precession of the neutron star, which may even be non-deformed, is caused by electromagnetic forces induced by stellar rotation. As a result, the internal magnetic field an order of magnitude less appears to be sufficient to explain the observed FRB periodicity.

\section{Precession, electric current, and corotation paradox}

A rotating body the ellipsoid of inertia of which is a sphere cannot precess freely (the sphere and cube of constant density are the simplest examples of such bodies). Free precession of a magnetized neutron star stems from the appearance of stellar deformation, e.g., caused by a sufficiently strong magnetic field. The deformed sphere loses spherical symmetry, and the tensor of inertia is no longer proportional to the unit matrix, so that one has a biaxial or triaxial situation leading to precession even without applied moment of forces.

However, if the ellipsoid of inertia remains a sphere, e.g., when magnetic deformation is ineffective because of insufficiently strong magnetic field or its complex structure leading to possible compensation, precession can appear if the moment of forces acting on the body is non-zero. This forced precession, once one assumes the hypothetic link between periodic FRB repeaters and precessing neutron stars, will have the same observational consequences as free precession, which are already discussed thoroughly in \citep{LevinBeloborodovBransgrove2020,ZanazziLai2020}. We will not repeat them here and refer the interested reader to these works but instead look into the origin of forced precession.

Analogously to \citep{ZanazziLai2020}, we model a neutron star as a perfectly conducting magnetized sphere that rigidly rotates about a fixed point (the form may be arbitrary). The internal electric and magnetic fields then satisfy \citep{Sobyanin2016}
\begin{gather}
\label{rotationBEq}
\frac{d\mathbf{B}}{dt}=\mathbf{\Omega}\times\mathbf{B},
\\
\label{rotationEForArbitraryOmega}
\frac{d\mathbf{E}}{dt}=\mathbf{\Omega}\times\mathbf{E}-\mathbf{w}\times\mathbf{B},
\end{gather}
where $d/dt=\partial/\partial t+\mathbf{v}\cdot\nabla$ is the full time derivative, $\mathbf{v}=\mathbf{\Omega}\times\mathbf{r}$ is the velocity, and $\mathbf{w}=\dot{\mathbf{\Omega}}\times\mathbf{r}$ is the rotational acceleration ($c=1$ throughout the paper). The rotation is assumed arbitrary, i.e., the absolute value and direction of the angular velocity $\mathbf{\Omega}$ may change with time. Equation \eqref{rotationBEq} shows magnetic corotation: the magnetic field is frozen-in in the medium and moves synchronously with it even when the angular velocity vector changes with time. By contrast, equation \eqref{rotationEForArbitraryOmega} indicates that the electric field corotates when the angular velocity vector is constant, in which case $\mathbf{w}=0$ and $d\mathbf{E}/dt=\mathbf{\Omega}\times\mathbf{E}$. The electric and magnetic fields are related via the condition of infinite conductivity,
\begin{equation}
\label{forceFree}
\mathbf{E}=-\mathbf{v}\times\mathbf{B}.
\end{equation}

The total electric current is determined by one of Maxwell's equations,
\begin{equation}
\label{j}
\mathbf{j}=\frac{1}{4\upi}\biggl(\nabla\times\mathbf{B}-\frac{\partial\mathbf{E}}{\partial t}\biggr),
\end{equation}
and can conveniently be represented as the sum of three parts,
\begin{equation}
\label{totalCurrent}
\mathbf{j}=\mathbf{j}_m+\mathbf{j}_{\Omega}+\mathbf{j}_{\dot{\Omega}}.
\end{equation}
where
\begin{equation}
\label{magnetizationCurrent}
\mathbf{j}_m=\frac{\nabla\times\mathbf{B}}{4\upi},
\end{equation}
is the magnetization current existing irrespective of the values of $\mathbf{\Omega}$ and $\dot{\mathbf{\Omega}}$,
\begin{equation}
\label{rotationCurrentViaRot}
\mathbf{j}_{\Omega}=\frac{\mathbf{v}}{4\upi}\times\bigl[\nabla\times(\mathbf{v}\times\mathbf{B})\bigr].
\end{equation}
is the rotation current existing when $\mathbf{\Omega}\neq0$, and
\begin{equation}
\label{accelerationCurrent}
\mathbf{j}_{\dot{\Omega}}=\frac{\mathbf{w}\times\mathbf{B}}{4\upi}.
\end{equation}
is the acceleration current existing when $\dot{\mathbf{\Omega}}\neq0$.

Now consider a sphere rotating with constant angular velocity. The acceleration current then vanishes and the total current is the sum of the magnetization and rotation currents, $\mathbf{j}=\mathbf{j}_m+\mathbf{j}_{\Omega}$. Let us examine whether the current corotates. The magnetization current corotates always: since the magnetic field is frozen in, the magnetization current, proportional to its curl, is also frozen in, and for $\mathbf{j}_m$ the same corotation equation as for $\mathbf{B}$ is valid, $d\mathbf{j}_m/dt=\mathbf{\Omega}\times\mathbf{j}_m$. When $\mathbf{\Omega}$ is constant, the total acceleration is the axipetal acceleration only, $d\mathbf{v}/d t=\mathbf{\Omega}\times\mathbf{v}$, which implies corotation of the velocity; therefore, the rotation current corotates as it is expressed via the corotating quantities $\mathbf{v}$ and~$\mathbf{B}$, see~\eqref{rotationCurrentViaRot}. We see that the total current density corotates for constant~$\mathbf{\Omega}$,
\begin{equation}
\label{rotationJ}
\frac{d\mathbf{j}}{dt}=\mathbf{\Omega}\times\mathbf{j}.
\end{equation}

Since the internal electric field also corotates for constant~$\mathbf{\Omega}$, so does the charge density, determined by its divergence, $\rho=\nabla\cdot\mathbf{E}/4\upi$:
\begin{equation}
\label{rhoCorotation}
\frac{d\rho}{d t}=0.
\end{equation}
The charge density at every point in the sphere moves with the velocity of the medium at the point, so it seems natural that the current is the so-called corotation current \citep{BeskinZheltoukhov2014}
\begin{equation}
\label{jc}
\mathbf{j}_c=\rho\mathbf{v},
\end{equation}
which could appear due to the motion of the corotating charge density, i.e., that the charge and current densities are connected via the corotation relation $\mathbf{j}=\mathbf{j}_c$.

Remarkably, this is not so,
\begin{equation}
\label{jNotRhoV}
\mathbf{j}\neq\mathbf{j}_c,
\end{equation}
which constitutes a corotation paradox. In everyday situation, when we have a resting conducting matter subject to external action of electromotive force (say, a wire with an electric current generated by an external battery), the paradox does not emerge because the current is created by an external electric field, not by the motion of the matter itself, and is determined by the resistance. Meanwhile, in the framework of ideal relativistic magnetohydrodynamics, when any external electromagnetic action is absent and the charges and currents are related to the motion of the perfectly conducting medium having zero electric field in the comoving frame of reference, the question as to whether $\mathbf{j}=\mathbf{j}_\text{c}$ becomes non-trivial. For example, when relativistic motion of a plasma in the M87 jet is considered, the charge density (times~$c$) per unit length in a magnetic tube appears to be in the order of the electric current flowing in the tube without any artificial assumptions \citep{Sobyanin2017}. However, we will see on a specific example that were the corotation relation to hold generally, Maxwell's equations would not be satisfied.

\section{Illustrative example}

In relation to the problem of the anomalous torque, \citet{BeskinZheltoukhov2014} discuss an example: A neutron star of radius $R$ has a uniform magnetic field inside a sphere of radius $R_0<R$, the centre of which coincides with the stellar centre. Outside the sphere, inside the area $R_0<r<R$, the neutron star has a dipole magnetic field. The magnetic field distribution is not smooth on the sphere $r=R_0$. On the basis of this distribution we will construct a magnetic field distribution that has no singularities and is smooth, infinitely differentiable, everywhere in the neutron star. Such a distribution will allow us to have smooth and finite charge and current densities at every point, avoid singularity problems, and demonstrate explicitly violation of the corotation relation.

We consider a conducting sphere of radius $R$ in which there are three distinct areas: an inner sphere $0\leqslant r<R_1$, an intermediate spherical layer $R_1\leqslant r<R_2$, and an outer spherical layer $R_2\leqslant r<R$. We assume that the magnetic field is uniform in the area $0\leqslant r<R_1$ and dipole in the area $R_2\leqslant r<R$. In the intermediate area $R_1\leqslant r<R_2$ the magnetic field should somehow connect smoothly the fields in the inner and outer areas.

Let us construct such a smooth field. It can be represented via a vector potential~$\mathbf{A}$, like any magnetic field,
\begin{equation}
\label{BRotA}
\mathbf{B}=\nabla\times\mathbf{A}.
\end{equation}
The vector potential $\mathbf{A}=\mathbf{B}_0\times\mathbf{r}/2$ corresponds to the uniform field $\mathbf{B}_0$ in the inner area, and the vector potential $\mathbf{A}=\mathbf{m}\times\mathbf{r}/r^3$ corresponds to the the dipole field in the outer area, where $\mathbf{m}=m\mathbf{e}_m$ is the magnetic dipole moment parallel to $\mathbf{B}_0=B_0\mathbf{e}_m$ and giving the direction of the magnetic axis $\mathbf{e}_m$. We then need to construct a vector potential for the intermediate area.

Firstly, note that we may consider the field $\mathbf{B}_0$ in the inner area as the field at the magnetic axis. Secondly, the field at the magnetic axis in the outer area is $2\mathbf{m}/r^3$, and we can present the vector potential in the form analogous to the form of the vector potential in the inner area, $\mathbf{A}=(2\mathbf{m}/r^3)\times\mathbf{r}/2$. Thus, in the case of the absence of the intermediate area, when $R_0=R_1=R_2$, we can write the overall vector potential for the whole area $0\leqslant r<R$ in the form
\begin{equation}
\label{AViaBm}
\mathbf{A}=\frac{\mathbf{B}_m\times\mathbf{r}}{2},
\end{equation}
where
\begin{equation}
\label{BmVector}
\mathbf{B}_m=B_m(r)\,\mathbf{e}_m
\end{equation}
is the magnetic field at the magnetic axis, whose absolute value $B_m(r)$ depends on the radial coordinate only. Since on the sphere $r=R_0$ the normal component of $\mathbf{B}$ is continuous due to $\nabla\cdot\mathbf{B}=0$, the function $B_m(r)$ is also continuous, though not smooth: $B_m(r)=B_0$ when $0\leqslant r<R_0$ and $B_n(r)=B_0(R_0/r)^3$ when $R_0\leqslant r<R$.

In the general case $R_1<R_2$, when the intermediate area exists, we construct the vector potential so that it has the same form \eqref{AViaBm} in the whole area $0\leqslant r<R$ and the absolute value of the field at the magnetic axis is smooth and coincides with $B_m(r)$ for the uniform field in the inner area and for the dipole field in the outer area. The vector potential is then smooth, and so is the magnetic field~\eqref{BRotA}.

To make a transition from the uniform to dipole field in the intermediate area $R_1\leqslant r<R_2$, we choose the total field as a weighted sum of the uniform and dipole fields, $B_m(r)=[1-F(r)]B_0+F(r)B_0(R_1/r)^3$, where $F(r)$ is a weight corresponding to the proportion of the dipole field in the whole field. The weight is $0$ in the inner and $1$ in the outer area, and it increases monotonously from $0$ to $1$ as $r$ increases from $R_1$ to $R_2$. We should choose a smooth $F(r)$ for a smooth $B_m(r)$. We may consider $F(r)$ a probability distribution function, and the corresponding probability density $p(r)=F'(r)$ is smooth and vanishes for $r\leqslant R_1$ and $r\geqslant R_2$. The $F(r)$ itself is then $\int^r p(x)dx$.

It remains to choose $p(x)$. We start with a smooth cap function \citep{Vladimirov2002}
\begin{equation}
\label{capFunction}
\omega_\varepsilon(x)=
\begin{cases}
\dfrac{C_0}{\varepsilon}\exp\biggl(-\dfrac{\varepsilon^2}{\varepsilon^2-x^2}\biggr),\quad -\varepsilon<x<\varepsilon,\\
0,\quad x\leqslant-\varepsilon\text{ or }x\geqslant\varepsilon,
\end{cases}
\end{equation}
where $\varepsilon>0$ and $C_0=1/\int_{-1}^1\exp[-1/(1-x^2)]dx\approx2.25$, so that $\omega_\varepsilon(x)$ is properly normalized, $\int\omega_\varepsilon(x)dx=1$. We can construct $p(x)$ with the necessary properties thus: $p(x)=\omega_\varepsilon(x-R_0)$, where $R_0=(R_1+R_2)/2$ and $\varepsilon=(R_2-R_1)/2$. Then we find $F(r)$ and obtain the smooth field at the magnetic axis,
\begin{equation}
\label{smoothBm}
B_m(r)=
\begin{cases}
B_0,\text{ }0\leqslant r<R_1,\\
B_0\int_{r-R_0}^\varepsilon\omega_\varepsilon(x)dx+B_0(R_1/r)^3\int_{-\varepsilon}^{r-R_0}\omega_\varepsilon(x)dx,\\R_1\leqslant r<R_2,\\
B_0(R_1/r)^3,\text{ }R_2\leqslant r<R.
\end{cases}
\end{equation}

Finally, we calculate the smooth magnetic field in the whole sphere by using \eqref{BRotA}--\eqref{BmVector},
\begin{equation}
\label{generalB}
\mathbf{B}=\biggl[B_m(r)+\frac{1}{2}\,r B'_m(r)\biggr]\mathbf{e}_m
-\frac{1}{2}\,r B'_m(r)\,\mathbf{e}_r\mathbf{e}_r\cdot\mathbf{e}_m,
\end{equation}
where prime denotes the $r$ derivative and $\mathbf{e}_r=\mathbf{r}/r$.

Let us find the form of the magnetic field lines. We calculate from \eqref{generalB} the components of the magnetic field in the spherical coordinates $r$, $\theta$, $\phi$ with the magnetic axis $\mathbf{e}_m$ as the polar axis:
\begin{gather}
\label{Br}
B_r=B_m(r)\cos\theta,
\\
\label{Btheta}
B_\theta=-\biggl[B_m(r)+\frac{1}{2}\,r B'_m(r)\biggr]\sin\theta,
\\
\label{Bphi}
B_\phi=0.
\end{gather}
The differential equation for a magnetic field line is
\begin{equation}
\label{fieldLineDifEq}
\frac{dr}{B_r}=\frac{rd\theta}{B_\theta}
\end{equation}
and means that an elementary line element $d\mathbf{l}=(dr,rd\theta)$ is parallel to the field. Noticing that $B_m(r)+r B'_m(r)/2=[r^2B_m(r)]'/2r$, we integrate \eqref{fieldLineDifEq} and arrive at the equation determining the form of the magnetic field lines,
\begin{equation}
\label{fieldLineEq}
(r\sin\theta)^2B_m(r)=C,
\end{equation}
where $C$ is a constant parameterizing the lines and being related to their position in the sphere. In the inner area equation \eqref{fieldLineEq} reduces to $r\sin\theta=r_1$ and describes a line parallel to the magnetic axis, with $r_1=\sqrt{C/B_0}$ being the distance between the line and the axis. In the outer area \eqref{fieldLineEq} takes on the form $r=r_2(\sin\theta)^2$ and describes a dipole field line, with $r_2=B_0R_1^3/C$ being the distance between the centre and the point of intersection of the line with the magnetic equatorial plane $\theta=\upi/2$ in the outer area.

\begin{figure}
\centering{\includegraphics[width=0.7\columnwidth]{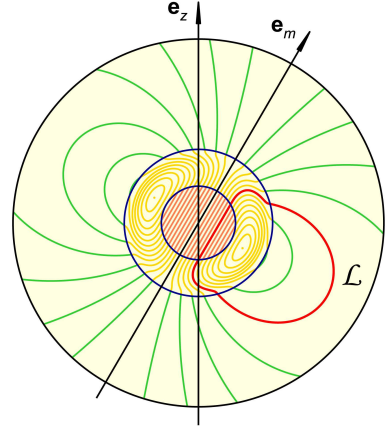}}
\caption{Illustrative smooth distribution of the magnetic field in the rotating conducting sphere.}
\label{fig1}
\end{figure}

Figure 1 shows the constructed smooth distribution of the internal magnetic field. We have chosen a sphere of radius $R$ with parameters $R_1=R/5$ and $R_2=2R/5$ in order to clearly show the field peculiarities. The magnetic field is uniform in the inner area $0\leqslant r<R_1$, dipole in the outer area $R_2\leqslant r<R$, and gives the required smooth transition in the intermediate area $R_1\leqslant r<R_2$. The magnetic field lines shown correspond to a series of $\sqrt{C/B_0}/R$ from $0$ (magnetic axis) to $0.24$ (last closed lines around the two points in the intermediate area) by step~$0.02$.  Some of the lines starting in the inner area close in the intermediate area, the rest pass through the intermediate area and either close in the outer area or go outside the sphere. Some closed lines lie entirely in the intermediate area. The lines degenerate into the points when the magnetic field vanishes, in which case $\sqrt{C/B_0}/R$ reaches a maximum of $\approx0.25$. The points lie in the magnetic equatorial plane at a distance of $\approx0.27R$ from the magnetic axis.

\section{Violation of Maxwell's equations}

Now we may return to the corotation paradox and demonstrate that the relation $\mathbf{j}=\mathbf{j}_\text{c}$ does not hold generally and is in conflict with Maxwell's equations in our illustrative example. For a constant angular velocity, we get from Maxwell's equations
\begin{equation}
\label{rotBMinusVE}
\nabla\times(\mathbf{B}-\mathbf{v}\times\mathbf{E})=4\upi(\mathbf{j}-\mathbf{j}_\text{c}).
\end{equation}
Were $\mathbf{j}=\mathbf{j}_\text{c}$, we would have $\nabla\times(\mathbf{B}-\mathbf{v}\times\mathbf{E})=0$ and
\begin{equation}
\label{intRotBMinusVE}
\oint\bigl[(1-v^2)\mathbf{B}+\mathbf{v}\mathbf{v}\cdot\mathbf{B}\bigr]\cdot d\mathbf{l}=0,
\end{equation}
where integration is performed over arbitrary closed contour lying inside the sphere; we have used Stokes's formula and taken account of~\eqref{forceFree}.

The internal magnetic field is determined by~\eqref{generalB}, and the magnetic axis $\mathbf{e}_m$ is generally inclined with respect to the rotation axis~$\mathbf{e}_z$ ($\upi/6$ arbitrarily chosen in Fig.~\ref{fig1}). We may choose a contour of integration that lies in the plane passing through $\mathbf{e}_z$ and~$\mathbf{e}_m$. The integral of the last term in~\eqref{intRotBMinusVE} then vanishes because $\mathbf{B}$ is orthogonal to $\mathbf{v}$ in this plane,
\begin{equation}
\label{intZero}
\oint d\mathbf{l}\cdot\mathbf{v}\mathbf{v}\cdot\mathbf{B}=0.
\end{equation}
Among all contours lying in the plane we now choose a contour that coincides with a closed magnetic field line. For instance, we may choose as a contour the magnetic field line $\mathcal{L}$ given by \eqref{fieldLineEq} with $C=B_0R^2/100$ (see Fig.~\ref{fig1}). Since $v<1$, $\mathbf{B}\cdot d\mathbf{l}=B\,dl$, and $B>0$, we have
\begin{equation}
\label{intOneMinusVSquareB}
\oint(1-v^2)\mathbf{B}\cdot d\mathbf{l}>0.
\end{equation}
Thus, equation \eqref{intRotBMinusVE} is clearly violated for the contour $\mathcal{L}$, which makes inequality \eqref{jNotRhoV} evident.

How to resolve the corotation paradox? The charge density corotates with the sphere, but the corotation is a special case of the change. A change in the charge density is brought about by the currents, and we have the charge conservation law
\begin{equation}
\label{chargeConservation}
\frac{\partial\rho}{\partial t}+\nabla\cdot\mathbf{j}=0.
\end{equation}
The currents are determined by the motion of different types of charges, but corotation of the charged particles themselves in not required to provide corotation of the charge density. The particles can move in any way if only such currents are provided whose divergence gives such a change in the charge density that a toroidal motion of this density with velocity of motion of the matter effectively takes place.

For completeness, consider the distribution of the electric current in our example. The total current contains two contributions,
\begin{equation}
\label{jmExample}
\mathbf{j}_m=\bigl[4B'_m(r)+rB''_m(r)\bigr]\frac{\mathbf{e}_r\times\mathbf{e}_m}{8\upi}
\end{equation}
and
\begin{equation}
\label{jOmegaExample}
\mathbf{j}_\Omega=\frac{\mathbf{e}_m\cdot\mathbf{v}}{4\upi}\biggl[B_m(r)\mathbf{\Omega}+\frac12 rB'_m(r)\mathbf{e}_r\mathbf{e}_r\cdot\mathbf{\Omega}\biggr].
\end{equation}

\begin{figure}
\centering{\includegraphics[width=0.7\columnwidth]{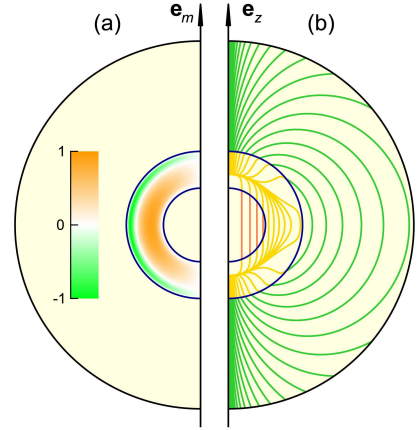}}
\caption{Illustrative distribution of the magnetization (a) and rotation (b) currents in the rotating conducting sphere.}
\label{fig2}
\end{figure}

The magnetization current $\mathbf{j}_m$ is a toroidal current symmetrical with respect to the magnetic axis $\mathbf{e}_m$ and is non-zero only in the intermediate area of the sphere. Figure \ref{fig2}(a) shows the $\mathbf{j}_m$ distribution in a half plane passing through~$\mathbf{e}_m$. The distribution is axisymmetric, i.e., does not depend on the azimuthal angle $\phi$ determining the position of the half plane. The magnetization current is orthogonal to the half plane, and its density is shown with the help of color scale. There are both the direct magnetization current whose direction agrees with the magnetic axis direction via the screw rule, and the inverse magnetization current whose direction is opposite. The positive density corresponds to the direct and the negative to the inverse magnetization current. The current density is normalized over the maximum inverse magnetization current, and the maximum direct magnetization current has a normalized density of $\approx0.90$.

The rotation current $\mathbf{j}_\Omega$, in contrast to $\mathbf{j}_m$, is a poloidal current, with the current lines lying in the planes passing through the rotation axis $\mathbf{e}_z$. Figure \ref{fig2}(b) shows the $\mathbf{j}_\Omega$ distribution in a half plane passing through $\mathbf{e}_z$. The distribution is not axisymmetric as the density and direction of the rotation current depend on the azimuthal angle $\varphi$ determining the position of the half plane (the value $\varphi=0$ may be ascribed to the half plane in which the positive direction of the magnetic axis lies), but the structure of the distribution is in itself the same, and the shown current lines are completely identical in any half plane. Note also that $\mathbf{j}_\Omega$ vanishes in the plane passing through the rotation and magnetic axes, which corresponds to $\varphi=0$ and $\upi$. When $0<\varphi<\upi$, the current lines intersect the equatorial plane from top to bottom, and when $\upi<\varphi<2\upi$ -- from bottom to top.

Thus, the real electric current differs from the formal corotation current both in magnitude and in direction. This is especially evident in the inner and outer areas of the sphere, where the magnetization current vanishes and the total current is the poloidal rotation current, which is orthogonal to velocity. In turn, the toroidal corotation current is parallel to velocity, so the electric current is orthogonal to the corotation current in these areas. Here an effective toroidal motion of the charge density is provided by poloidal currents. The same can be taken in the intermediate area since the non-zero magnetization current does not contribute to the change in the charge density, $\nabla\cdot\mathbf{j}_m=0$.

We can observe some hydrodynamic analogy with the case of gravity waves on a fluid surface: the waves propagate over the surface with a positive velocity, whereas the fluid particles, whose motion essentially determines the existence and type of the waves, do not move with this velocity but circle uniformly round fixed points, so that their average velocity is zero \citep{LandauLifshitz1987}.

\section{Anomalous torque}

We have already mentioned the possibility of precession of a neutron star as a result of deformation, and then we dealt with free precession, when the torque, the moment of forces, is absent (Euler's case). Under the influence of a non-zero torque the neutron star experiences forced precession, and the torque is of electromagnetic nature,
\begin{equation}
\label{momentOfForces}
\mathbf{M}=\int\mathbf{r}\times(\rho\mathbf{E}+\mathbf{j}\times\mathbf{B})\,d V,
\end{equation}
where integration is performed over the stellar volume including the surface. The torque inducing forced precession emerges as a result of stellar rotation and represents the so-called anomalous torque, i.e., the torque significantly exceeding the usual spin-down torque, by the factor $(\Omega R)^{-1}\sim10^3-10^4$. Despite the long history of the subject \citep{DavisGoldstein1970,Goldreich1970,Melatos2000,Istomin2005,BeskinZheltoukhov2014,ZanazziLai2015}, so far there is no single opinion on how to calculate the torque and what is its exact value even in the case of a rotating uniformly magnetized sphere. Meanwhile, with all necessary fields, charges, and currents in hand, we are in position to calculate this torque using the direct method, i.e., summing the moments of all Lorentz forces acting on the sphere with equation~\eqref{momentOfForces}.

We consider the case of a neutron star with a plasma-filled magnetosphere. We assume that electromagnetic discontinuities at $r=R$ and corresponding surface charges and currents are absent because of the existence of a dense surrounding electron-positron plasma. Even if the standard pulsar-like plasma generation does not occur, the neutron star is immersed in the background of cosmic gamma radiation, which illuminates the magnetosphere and causes one-photon pair production giving the needed plasma \citep{IstominSobyanin2011a,IstominSobyanin2011b}.

The discussion of the validity of the corotation relation given in detail in the preceding sections allows us to conclude that $\mathbf{M}\neq0$. This is very important in light of the earlier conclusion that $\mathbf{M}=0$ based on the corotation relation, when a torque on the conducting sphere in vacuo can be generated by surface discontinuities only  \citep{BeskinZheltoukhov2014}. The latter conclusion can readily be verified by substituting $\mathbf{j}=\rho\mathbf{v}$ into \eqref{momentOfForces} and remembering \eqref{forceFree}. It also remains in force when the surface charges and currents are non-zero but related analogously by $\mathbf{j}_\text{surf}=\rho_\text{surf}\mathbf{v}$. We see that the corotation relation should never be adopted artificially. In our case, inequality \eqref{jNotRhoV} underlies the very possibility of considering forced precession as a probable reason for the observed periodic activity of FRBs.

For a uniform internal magnetic field $\mathbf{B}$, the magnetization current vanishes, so the total current is given by \eqref{jOmegaExample},
\begin{equation}
\label{current}
\mathbf{j}=\mathbf{\Omega}\frac{\mathbf{v}\cdot\mathbf{B}}{4\upi}=-\mathbf{\Omega}\mathbf{r}\cdot\frac{\mathbf{\Omega}\times\mathbf{B}}{4\upi},
\end{equation}
while the charge density is constant,
\begin{equation}
\label{charge}
\rho=-\frac{\mathbf{\Omega}\cdot\mathbf{B}}{2\upi}.
\end{equation}
The electric field has the form
\begin{equation}
\label{electric}
\mathbf{E}=\mathbf{\Omega}\mathbf{r}\cdot\mathbf{B}-\mathbf{r}\mathbf{\Omega}\cdot\mathbf{B}.
\end{equation}
Substituting equations \eqref{current}--\eqref{electric} into \eqref{momentOfForces} then gives
\begin{equation}
\label{momentOfForces1}
\mathbf{M}=\int\Bigl[-\rho\mathbf{\Omega}\times\mathbf{r}\mathbf{r}\cdot\mathbf{B}
+\frac{\mathbf{\Omega}\times\mathbf{B}}{4\upi}\times\mathbf{r}\mathbf{r}\cdot(\mathbf{\Omega}\times\mathbf{B})\Bigr]d V.
\end{equation}
Since
\begin{equation}
\label{intrrdv}
\int\mathbf{r}\mathbf{r}\,d V=\frac{4\upi}{15}R^5\mathbf{I},
\end{equation}
where $\mathbf{I}$ is the unit tensor, we use $\mathbf{a}\times\mathbf{I}\cdot\mathbf{b}=\mathbf{a}\times\mathbf{b}$ and finally get the anomalous electromagnetic torque acting on the rotating uniformly magnetized perfectly conducting sphere in the absence of surface charges and currents,
\begin{equation}
\label{volumeTorque}
\mathbf{M}=\frac{2R^5}{15c^2}\mathbf{\Omega}\times\mathbf{B}\mathbf{\Omega}\cdot\mathbf{B}=\frac8{15Rc^2}\mathbf{\Omega}\times\mathbf{m}\mathbf{\Omega}\cdot\mathbf{m},
\end{equation}
where $\mathbf{m}=\mathbf{B}R^3/2$ is the magnetic moment ($c$ restored). Evidently, $\mathbf{M}\neq0$. The value \eqref{volumeTorque} differs from previous results \citep{DavisGoldstein1970,Goldreich1970,Melatos2000,Istomin2005,BeskinZheltoukhov2014,ZanazziLai2015}, but there is no contradiction because only the vacuum torque was studied in these works.

\section{Magnetic field estimation for periodic FRBs}

It remains to relate the torque to the observed precessional period $P_\text{prec}$, presumably being the modulation period of the FRB activity, and estimate the corresponding internal magnetic field. The rotating neutron star is analogous to the gyroscope, and if we consider a biaxial gyroscope with principal moments of inertia $A=B$ and $C$, the latter corresponding to the axis of dynamical symmetry, the basic gyroscopic formula reads \citep{Markeev1999}
\begin{equation}
\label{basicGyroscopicFormula}
\mathbf{M}=\boldsymbol{\omega}_2\times\boldsymbol{\omega}_1\Bigl[C+(C-A)\frac{\omega_2}{\omega_1}\cos\theta_0\Bigr],
\end{equation}
where $\mathbf{M}$ is the torque inducing forced precession, $\boldsymbol{\omega}_1$ and $\boldsymbol{\omega}_2$ are the angular velocities of proper rotation and precession, respectively, and $\theta_0$ is the angle of nutation. Note that precession of the neutron star means rotation of the angular velocity around the axis of dynamical symmetry in the rotating frame of reference frozen-in into the star. Therefore, the angular velocity of precession of the neutron star, $\mathbf{\Omega}_\text{prec}$, corresponds to the angular velocity of proper rotation of the equivalent gyroscope, $\boldsymbol{\omega}_1=\mathbf{\Omega}_\text{prec}$. The angular velocities $\mathbf{\Omega}$ and $\mathbf{\Omega}_\text{prec}$ simultaneously execute precession in the laboratory frame of reference with angular velocity $\boldsymbol{\omega}_2=\mathbf{\Omega}-\mathbf{\Omega}_\text{prec}$.

In the case of an anomalous torque, the magnetic axis is considered as the precessional axis, $\mathbf{\Omega}_\text{prec}=\Omega_\text{prec}\mathbf{e}_m$, where $\Omega_\text{prec}=2\upi/P_\text{prec}$ is the precessional frequency of the neutron star. Then putting $\omega_2\approx\Omega$ and $\theta_0\approx\theta_m$ in \eqref{basicGyroscopicFormula} on the basis of the inequality $P_\text{prec}\gg P$, where $\Omega=2\upi/P$ is the rotational frequency, $P$ is the rotational period, and $\theta_m$ is the angle of magnetic inclination, and comparing to \eqref{volumeTorque}, we obtain the precessional period
\begin{equation}
\label{PprecViaEpsilon}
P_\text{prec}=\frac{P}{(\epsilon+\epsilon_\text{eff})\cos{\theta_m}},
\end{equation}
where $\epsilon=(A-C)/C$ is the parameter of deformation and
\begin{equation}
\label{effectiveDeformationParameter}
\begin{split}
\epsilon_\text{eff}&=\frac{2}{15c^2}\frac{R^5B^2}{C}\\
&=1.33\times10^{-7}\Bigl(\frac{M}{1.4 M_\odot}\Bigr)^{-1}\Bigl(\frac{R}{10\text{ km}}\Bigr)^3\Bigl(\frac{B}{10^{15}\text{ G}}\Bigr)^2,
\end{split}
\end{equation}
with $M$ being the stellar mass and $C\approx J=2MR^2/5$, is an effective parameter corresponding to a fictitious deformation such that in a fully dynamically symmetric case, when $\epsilon=0$, forced precession of the non-deformed star is equivalent to free precession of the so-deformed star.

\citet{ZanazziLai2020} adopt the different effective parameter $\epsilon_\text{eff}^*=3R^5B^2/20c^2J$, which will take place if one assumes that the anomalous torque is given by the flux of electromagnetic angular momentum entering the surface of the neutron star rotating in vacuo. On the one hand, this flux differs from the torque calculated by the direct method; besides, the vacuum magnetosphere is tacitly implied. On the other hand, our consistent consideration of electromagnetic fields, charges, and currents gives almost the same effective deformation parameter $\epsilon_\text{eff}\approx\epsilon_\text{eff}^*$, with the factor $1.33\times10^{-7}$ instead of $1.5\times10^{-7}$ in equation \eqref{effectiveDeformationParameter}. Thus, the results of this work justify the previous estimates of \citet{ZanazziLai2020} in the numerical sense and, most importantly, give the sense to consideration of the effect of anomalous torque on precession, implied to be zero in the case of corotational currents.

The anomalous torque in not the only reason for precession, and other contributions are related to the real stellar deformation inducing free precession. For example, the precessional frequencies from magnetic deformation $\epsilon_B$ of the stellar matter are expected to be larger than those from the anomalous torque by the ratio \citep{Melatos2000,ZanazziLai2015}
\begin{equation}
\label{epsilonRatio}
\frac{\epsilon_B}{\epsilon_\text{eff}}=\lambda\frac{R}{R_\text{g}},
\end{equation}
where $R_\text{g}=GM/c^2$ is the gravitational radius and the dimensionless factor $\lambda$ depends on the exact field geometry, and can be less if the neutron star has a significant toroidal magnetic field \citep{MastranoLaskyMelatos2013,MastranoSuvorovMelatos2015}. Other contributions are considered in \citet{ZanazziLai2020} and are less effective, so we will not repeat these results here. We obtain from equations \eqref{PprecViaEpsilon}--\eqref{epsilonRatio} the following estimate for the internal magnetic field of a precessing neutron star,
\begin{equation}
\label{Bint}
\begin{split}
B&=c\,\sqrt{\frac{15J}{2R^5\cos{\theta_m}}\frac{P}{P_\text{prec}}\frac1{1+\lambda R/R_\text{g}}}\\
&\sim7.45\times10^{17}\text{ G}\,\sqrt{\frac{P}{P_\text{prec}}},
\end{split}
\end{equation}
where $M=1.4M_\odot$, $R=10$~km, $\cos\theta_m=1$, and $\lambda=-3$ (corresponding to $\beta=-1$ in $\epsilon_B=\beta R^4 B^2/GM^2$) are adopted in the rightmost expression. The negative $\beta$ is chosen for formal consistency with using the poloidal field in the calculations of the anomalous torque, which induces oblateness, but the positive $\beta$ will give a comparable estimate.

The rotational period is unknown, and we may only assume the typical rotational period for neutron stars, $P\sim1$~s. For FRB 180916 with $P_\text{prec}=16.35$~d we then have $B\sim6\times10^{14}$~G, which is the magnetar field but is an order of magnitude less than the previously assumed $10^{16}$~G. If such situation is indeed realized, relation \eqref{RBtau} proposed by \citet{TavaniEtal2020} no longer excludes the magnetar nature of FRBs: for the maximum observational time $\tau\sim10^5$~s the admissable magnetic field is $\sim3\times10^{14}$~G; this constraint may simply imply slightly faster rotation of the possible progenitor of FRB 180916, with $P\sim0.2-0.3$~s. As to another possibly repeating FRB 121102 with $P_\text{prec}=157$~d, for $P\sim1$~s we have $B\sim2\times10^{14}$~G not exceeding the admissible field, so that $P\sim2-3$~s is possible. Note finally that the dissipative times probably exceed the observational times of \citet{TavaniEtal2020} and allow longer rotational periods and higher magnetic fields. Destruction of precession by vortex pinning in the obtained relatively low magnetic fields \citep{Shaham1977,LinkEpstein1997,SedrakianWassermanCordes1999,AkgunLinkWasserman2006} may be prevented by the assumption that the superfluid is pinned in the region located in the stellar interior \citep{GoglichidzeBarsukov2019}.

\section{Conclusions}

In this work the connection of the recently detected periodic FRB repeaters with magnetized neutron stars precessing in a forced manner has been studied. The study is motivated by trying to overcome the constraints on the magnetic field that follow from the joint X-ray and $\gamma$-ray observations of FRB 180916 and pose some difficulties for considerations of strongly magnetized neutron stars. Forced precession takes place even if stellar deformation is absent, but then it requires a torque acting on the star. Directly summing the moments of Lorentz forces, we have shown that this anomalous torque of electromagnetic nature appears to be non-zero and owes its existence to non-corotational electric currents induced by stellar rotation. Using the gyroscopic formula to consider the effects of both the torque and the deformation, we have estimated the internal magnetic field through the rotational and precessional periods of the neutron star, the latter period presumably corresponding to the observed periodic modulation of the repeating FRB activity. The obtained values appear to be an order of magnitude less than those bringing problems from the conclusions of the high-energy observations and thus allow the hypothesis of the origin of at least some FRBs from precessing neutron stars to remain plausible.

\section{Acknowledgements}

I thank the reviewer for valuable comments and suggestions.

\section{Data availability}

All data are incorporated into the article.

%%%%%%%%%%%%%%%%%%%%%%%%%%%%%%%%%%%%%%%%%%%%%%%%%%

%%%%%%%%%%%%%%%%%%%% REFERENCES %%%%%%%%%%%%%%%%%%

% The best way to enter references is to use BibTeX:

\bibliographystyle{mnras}

\begin{thebibliography}{}
\makeatletter
\relax
\def\mn@urlcharsother{\let\do\@makeother \do\$\do\&\do\#\do\^\do\_\do\%\do\~}
\def\mn@doi{\begingroup\mn@urlcharsother \@ifnextchar [ {\mn@doi@}
  {\mn@doi@[]}}
\def\mn@doi@[#1]#2{\def\@tempa{#1}\ifx\@tempa\@empty \href
  {http://dx.doi.org/#2} {doi:#2}\else \href {http://dx.doi.org/#2} {#1}\fi
  \endgroup}
\def\mn@eprint#1#2{\mn@eprint@#1:#2::\@nil}
\def\mn@eprint@arXiv#1{\href {http://arxiv.org/abs/#1} {{\tt arXiv:#1}}}
\def\mn@eprint@dblp#1{\href {http://dblp.uni-trier.de/rec/bibtex/#1.xml}
  {dblp:#1}}
\def\mn@eprint@#1:#2:#3:#4\@nil{\def\@tempa {#1}\def\@tempb {#2}\def\@tempc
  {#3}\ifx \@tempc \@empty \let \@tempc \@tempb \let \@tempb \@tempa \fi \ifx
  \@tempb \@empty \def\@tempb {arXiv}\fi \@ifundefined
  {mn@eprint@\@tempb}{\@tempb:\@tempc}{\expandafter \expandafter \csname
  mn@eprint@\@tempb\endcsname \expandafter{\@tempc}}}

\bibitem[\protect\citeauthoryear{{Akg\"{u}n}, Link  \& Wasserman}{{Akg\"{u}n}
  et~al.}{2006}]{AkgunLinkWasserman2006}
{Akg\"{u}n} T.,  Link B.,   Wasserman I.,  2006, MNRAS, 365, 653

\bibitem[\protect\citeauthoryear{Amiri et~al.}{Amiri
  et~al.}{2020}]{AmiriEtal2020}
Amiri M.,  et~al., 2020, Nature, 582, 351

\bibitem[\protect\citeauthoryear{Andersen et~al.}{Andersen
  et~al.}{2019}]{AndersenEtal2019}
Andersen B.~C.,  et~al., 2019, ApJ, 885, L24

\bibitem[\protect\citeauthoryear{Bannister et~al.}{Bannister
  et~al.}{2019}]{BannisterEtal2019}
Bannister K.~W.,  et~al., 2019, Science, 365, 565

\bibitem[\protect\citeauthoryear{Beniamini, Wadiasingh  \& Metzger}{Beniamini
  et~al.}{2020}]{BeniaminiWadiasinghMetzger2020}
Beniamini P.,  Wadiasingh Z.,   Metzger B.~D.,  2020, preprint
  (arXiv:2003.12509)

\bibitem[\protect\citeauthoryear{Beskin \& Zheltoukhov}{Beskin \&
  Zheltoukhov}{2014}]{BeskinZheltoukhov2014}
Beskin V.~S.,  Zheltoukhov A.~A.,  2014, Phys. Usp., 57, 799

\bibitem[\protect\citeauthoryear{Chawla et~al.}{Chawla
  et~al.}{2020}]{ChawlaEtal2020}
Chawla P.,  et~al., 2020, ApJ, 896, L41

\bibitem[\protect\citeauthoryear{Chernoff, Li  \& Tye}{Chernoff
  et~al.}{2020}]{ChernoffLiTye2020}
Chernoff D.~F.,  Li S.~Y.,   Tye S.-H.~H.,  2020, preprint (arXiv:2003.07275)

\bibitem[\protect\citeauthoryear{Connor, Miller  \& Gardenier}{Connor
  et~al.}{2020}]{ConnorMillerGardenier2020}
Connor L.,  Miller M.~C.,   Gardenier D.~W.,  2020, preprint (arXiv:2003.11930)

\bibitem[\protect\citeauthoryear{Cordes \& Chatterjee}{Cordes \&
  Chatterjee}{2019}]{CordesChatterjee2019}
Cordes J.~M.,  Chatterjee S.,  2019, ARA\&A, 57, 417

\bibitem[\protect\citeauthoryear{Dai \& Zhong}{Dai \&
  Zhong}{2020}]{DaiZhong2020}
Dai Z.~G.,  Zhong S.~Q.,  2020, ApJ, 895, L1

\bibitem[\protect\citeauthoryear{Davis \& Goldstein}{Davis \&
  Goldstein}{1970}]{DavisGoldstein1970}
Davis L.,  Goldstein M.,  1970, ApJ, 159, L81

\bibitem[\protect\citeauthoryear{Fonseca et~al.}{Fonseca
  et~al.}{2020}]{FonsecaEtal2020}
Fonseca E.,  et~al., 2020, ApJ, 891, L6

\bibitem[\protect\citeauthoryear{Goglichidze \& Barsukov}{Goglichidze \&
  Barsukov}{2019}]{GoglichidzeBarsukov2019}
Goglichidze O.~A.,  Barsukov D.~P.,  2019, MNRAS, 482, 3032

\bibitem[\protect\citeauthoryear{Goldreich}{Goldreich}{1970}]{Goldreich1970}
Goldreich P.,  1970, ApJ, 160, L11

\bibitem[\protect\citeauthoryear{Ioka \& Zhang}{Ioka \&
  Zhang}{2020}]{IokaZhang2020}
Ioka K.,  Zhang B.,  2020, ApJ, 893, L26

\bibitem[\protect\citeauthoryear{Istomin}{Istomin}{2005}]{Istomin2005}
Istomin {\relax Ya}.~N.,  2005, Magnetodipole Oven.
Nova Science, New York, pp 27--43

\bibitem[\protect\citeauthoryear{Istomin}{Istomin}{2018}]{Istomin2018}
Istomin {\relax Ya}.~N.,  2018, MNRAS, 478, 4348

\bibitem[\protect\citeauthoryear{Istomin \& Sob'yanin}{Istomin \&
  Sob'yanin}{2011a}]{IstominSobyanin2011a}
Istomin {\relax Ya}.~N.,  Sob'yanin D.~N.,  2011a, J. Exp. Theor. Phys., 113,
  592

\bibitem[\protect\citeauthoryear{Istomin \& Sob'yanin}{Istomin \&
  Sob'yanin}{2011b}]{IstominSobyanin2011b}
Istomin {\relax Ya}.~N.,  Sob'yanin D.~N.,  2011b, J. Exp. Theor. Phys., 113,
  605

\bibitem[\protect\citeauthoryear{Katz}{Katz}{2020}]{Katz2020}
Katz J.~I.,  2020, MNRAS, 494, L64

\bibitem[\protect\citeauthoryear{Kumar et~al.}{Kumar
  et~al.}{2019}]{KumarEtal2019}
Kumar P.,  et~al., 2019, ApJ, 887, L30

\bibitem[\protect\citeauthoryear{Landau \& Lifshitz}{Landau \&
  Lifshitz}{1987}]{LandauLifshitz1987}
Landau L.~D.,  Lifshitz E.~M.,  1987, Fluid Mechanics, second edn.
Pergamon, Oxford

\bibitem[\protect\citeauthoryear{Levin, Beloborodov  \& Bransgrove}{Levin
  et~al.}{2020}]{LevinBeloborodovBransgrove2020}
Levin Y.,  Beloborodov A.~M.,   Bransgrove A.,  2020, ApJ, 895, L30

\bibitem[\protect\citeauthoryear{Liebling \& Palenzuela}{Liebling \&
  Palenzuela}{2016}]{LieblingPalenzuela2016}
Liebling S.~L.,  Palenzuela C.,  2016, Phys. Rev. D, 6, 064046

\bibitem[\protect\citeauthoryear{Link \& Epstein}{Link \&
  Epstein}{1997}]{LinkEpstein1997}
Link B.,  Epstein R.~I.,  1997, ApJ, 478, L91

\bibitem[\protect\citeauthoryear{Lyutikov \& Popov}{Lyutikov \&
  Popov}{2020}]{LyutikovPopov2020}
Lyutikov M.,  Popov S.,  2020, preprint (arXiv:2005.05093)

\bibitem[\protect\citeauthoryear{Lyutikov, Barkov  \& Giannios}{Lyutikov
  et~al.}{2020}]{LyutikovBarkovGiannios2020}
Lyutikov M.,  Barkov M.~V.,   Giannios D.,  2020, ApJ, 893, L39

\bibitem[\protect\citeauthoryear{Makishima, Enoto, Hiraga, Nakano, Nakazawa,
  Sakurai, Sasano  \& Murakami}{Makishima et~al.}{2014}]{MakishimaEtal2014}
Makishima K.,  Enoto T.,  Hiraga J.~S.,  Nakano T.,  Nakazawa K.,  Sakurai S.,
  Sasano M.,   Murakami H.,  2014, Phys. Rev. Lett., 112, 171102

\bibitem[\protect\citeauthoryear{Marcote et~al.}{Marcote
  et~al.}{2020}]{MarcoteEtal2020}
Marcote B.,  et~al., 2020, Nature, 577, 190

\bibitem[\protect\citeauthoryear{Markeev}{Markeev}{1999}]{Markeev1999}
Markeev A.~P.,  1999, Theoretical Mechanics, second edn.
Regular and Chaotic Dynamics, Izhevsk

\bibitem[\protect\citeauthoryear{Mastrano, Lasky  \& Melatos}{Mastrano
  et~al.}{2013}]{MastranoLaskyMelatos2013}
Mastrano A.,  Lasky P.~D.,   Melatos A.,  2013, MNRAS, 434, 1658

\bibitem[\protect\citeauthoryear{Mastrano, Suvorov  \& Melatos}{Mastrano
  et~al.}{2015}]{MastranoSuvorovMelatos2015}
Mastrano A.,  Suvorov A.~G.,   Melatos A.,  2015, MNRAS, 447, 3475

\bibitem[\protect\citeauthoryear{Melatos}{Melatos}{2000}]{Melatos2000}
Melatos A.,  2000, MNRAS, 313, 217

\bibitem[\protect\citeauthoryear{Metzger, Berger  \& Margalit}{Metzger
  et~al.}{2017}]{MetzgerBergerMargalit2017}
Metzger B.~D.,  Berger E.,   Margalit B.,  2017, ApJ, 841, 14

\bibitem[\protect\citeauthoryear{Murase, Kashiyama  \& M\'{e}sz\'{a}ros}{Murase
  et~al.}{2016}]{MuraseKashiyamaMeszaros2016}
Murase K.,  Kashiyama K.,   M\'{e}sz\'{a}ros K.,  2016, MNRAS, 461, 1498

\bibitem[\protect\citeauthoryear{Pen \& Connor}{Pen \&
  Connor}{2015}]{PenConnor2015}
Pen U.-L.,  Connor L.,  2015, ApJ, 807, 179

\bibitem[\protect\citeauthoryear{Petroff, Hessels  \& Lorimer}{Petroff
  et~al.}{2019}]{PetroffHesselsLorimer2019}
Petroff E.,  Hessels J. W.~T.,   Lorimer D.~R.,  2019, A\&ARv, 27, 4

\bibitem[\protect\citeauthoryear{Pilia et~al.}{Pilia
  et~al.}{2020}]{PiliaEtal2020}
Pilia M.,  et~al., 2020, ApJ, 896, L40

\bibitem[\protect\citeauthoryear{Popov \& Postnov}{Popov \&
  Postnov}{2013}]{PopovPostnov2013}
Popov S.~B.,  Postnov K.~A.,  2013, preprint (arXiv:1307.4924)

\bibitem[\protect\citeauthoryear{Prochaska et~al.}{Prochaska
  et~al.}{2019}]{ProchaskaEtal2019}
Prochaska J.~X.,  et~al., 2019, Science, 366, 231

\bibitem[\protect\citeauthoryear{Rajwade et~al.}{Rajwade
  et~al.}{2020}]{RajwadeEtal2020}
Rajwade K.~M.,  et~al., 2020, MNRAS, 495, 3551

\bibitem[\protect\citeauthoryear{Ravi et~al.}{Ravi et~al.}{2019}]{RaviEtal2019}
Ravi M.,  et~al., 2019, Nature, 572, 352

\bibitem[\protect\citeauthoryear{Sedrakian, Wasserman  \& Cordes}{Sedrakian
  et~al.}{1999}]{SedrakianWassermanCordes1999}
Sedrakian A.,  Wasserman I.,   Cordes J.~M.,  1999, ApJ, 524, 341

\bibitem[\protect\citeauthoryear{Shaham}{Shaham}{1977}]{Shaham1977}
Shaham J.,  1977, ApJ, 214, 251

\bibitem[\protect\citeauthoryear{Sob'yanin}{Sob'yanin}{2016}]{Sobyanin2016}
Sob'yanin D.~N.,  2016, Astron. Lett., 42, 745

\bibitem[\protect\citeauthoryear{Sob'yanin}{Sob'yanin}{2017}]{Sobyanin2017}
Sob'yanin D.~N.,  2017, MNRAS, 471, 4121

\bibitem[\protect\citeauthoryear{Sob'yanin}{Sob'yanin}{2018}]{Sobyanin2018}
Sob'yanin D.~N.,  2018, MNRAS, 479, L65

\bibitem[\protect\citeauthoryear{Spitler et~al.}{Spitler
  et~al.}{2016}]{SpitlerEtal2016}
Spitler L.~G.,  et~al., 2016, Nature, 531, 202

\bibitem[\protect\citeauthoryear{Tavani et~al.}{Tavani
  et~al.}{2020}]{TavaniEtal2020}
Tavani M.,  et~al., 2020, ApJ, 893, L42

\bibitem[\protect\citeauthoryear{Tendulkar et~al.}{Tendulkar
  et~al.}{2017}]{TendulkarEtal2017}
Tendulkar S.~P.,  et~al., 2017, ApJ, 834, L7

\bibitem[\protect\citeauthoryear{Vladimirov}{Vladimirov}{2002}]{Vladimirov2002}
Vladimirov V.~S.,  2002, Methods of the Theory of Generalized Functions.
 Analytical Methods and Special Functions Vol. 6, Taylor \& Francis, London

\bibitem[\protect\citeauthoryear{Yang \& Dai}{Yang \& Dai}{2019}]{YangDai2019}
Yang Y.-H.,  Dai Z.-G.,  2019, ApJ, 885, 149

\bibitem[\protect\citeauthoryear{Yang \& Zou}{Yang \& Zou}{2020}]{YangZou2020}
Yang H.,  Zou Y.-C.,  2020, ApJ, 893, L31

\bibitem[\protect\citeauthoryear{Zanazzi \& Lai}{Zanazzi \&
  Lai}{2015}]{ZanazziLai2015}
Zanazzi J.~J.,  Lai D.,  2015, MNRAS, 451, 695

\bibitem[\protect\citeauthoryear{Zanazzi \& Lai}{Zanazzi \&
  Lai}{2020}]{ZanazziLai2020}
Zanazzi J.~J.,  Lai D.,  2020, ApJ, 892, L15

\makeatother
\end{thebibliography}

\providecommand{\noopsort}[1]{}\providecommand{\singleletter}[1]{#1}%

%%%%%%%%%%%%%%%%%%%%%%%%%%%%%%%%%%%%%%%%%%%%%%%%%%

% Don't change these lines
+\bsp	% typesetting comment
\label{lastpage}
\end{document}